\documentstyle[aps,preprint]{revtex}
\def\sb{\nabla{\hspace{-8pt}{\slash}}}
\begin{document} 
\draft 
\preprint{{IMSc-99/03/07};{CGPG-99/7/1};
{hep-th/9907116}} 
\title{Black Hole Emission Rates and the AdS/CFT Correspondence} 
\author{Saurya Das$^{1}$ \footnote{E-mail: das@gravity.phys.psu.edu}and,
 Arundhati Dasgupta$^2$ \footnote{E-mail: dasgupta@imsc.ernet.in}} 
\address{$^1$Center for Gravitational Physics and Geometry \\
Department of Physics, The Pennsylvania State University \\
University Park, PA 16802, USA \\
$^2$~The Institute of Mathematical Sciences,\\ CIT Campus, Taramani\\ 
Chennai - 600113, India \\}
\maketitle

\begin{abstract}
We study the emission rates of scalar, spinor and
vector particles from a 5 dimensional
black hole for arbitrary partial waves. The
solution is lifted to 6 dimensions, and the
near horizon $ BTZ \times S^3$ geometry of the
black hole solution is probed
to determine the greybody factors. We show that
the exact decay rates can be reproduced from a
$(1+1)$-dimensional conformal field theory 
which lies on the boundary of the near horizon geometry.
The AdS/CFT correspondence is used to determine
the dimension of the CFT operators corresponding
to the bulk fields. These operators couple
to plane waves incident on the CFT from infinity
to produce emission in the bulk.

\end{abstract}

\section{Introduction}
The study of the near horizon geometry of certain black holes 
has led to a better understanding of the origin of entropy
and Hawking radiation from an underlying conformal field theory 
\cite{malda1,strom1,spe,cl,teo}. The near horizon
geometry corresponds to anti-de Sitter
space with or without certain identifications and is
associated with a conformal field theory which lives
on 
the boundary of the anti-de Sitter space. The CFT obtained
thus carries non-trivial information about the 
black hole space-time. In
this paper we concentrate on studying emission
rates of particles
from 
a five dimensional
black hole and give a derivation
of the rates using the conformal field theory which is associated with
its near horizon geometry. The black hole
solution is obtained by
compactifying Type II B  string theory  
on $T^4 \times S^1$. On retaining the $S^1$ as a compact direction with a 
large radius, it gives a black string solution wrapped around the 
$S^1$. 
The near horizon geometry
of this configuration is $BTZ \times S^3$ where the $BTZ$ black hole 
is $3$-dimensional anti-de Sitter space with
certain identifications \cite{btz}. 
We observe that the emission rates of neutral particles 
obtained in the black string background are the same as that from the 
$5$-dimensional black hole \cite{malda1,sato}, and the near horizon
$BTZ$ geometry has a crucial role in determining the greybody factors
\cite{sid,adg,ohta1}. 
We thus study the matter fields obtained as
perturbations of the given $6$-dimensional supergravity background and
obtain the equations of motion of particles in the near horizon geometry, 
by considering a $AdS_3 \times S^3$ compactification of the six dimensional
supergravity. 
Since $BTZ$ space is locally
$AdS_3$, to study the equation of motion of
particles it suffices to study $AdS_3 \times
S^3$ compactification. The particles in 6 dimension are expanded in terms of the
harmonic functions on $S^3$, and higher partial wave objects
appear as massive excitations on the BTZ spacetime. We separately study
the behaviour of scalar, fermion
and vector particles of the  $6$-dimensional $N=8$ SUGRA spectrum. 
The fermions and vectors considered here are non-minimally coupled in
six dimensions . We look at arbitrary partial
waves for the particles and the greybody factors
are calculated by studying the wave equations in the near horizon
geometry and matching it suitably with the
wavefunctions in asymptotically flat spacetime
at a distance $r\sim l$ from the horizon, where $l$ is the $AdS_3$ radius.

The equation of motion of a minimally coupled scalar near the
black hole horizon reduces to the equation of motion of a 
massive scalar field in the BTZ background. By solving this
equation of motion, as well as the equation of motion in six dimensions, far from the
horizon, we match the
wavefunctions at an intermediate region
and 
determine the greybody factor. The latter agrees with the 
greybody factor obtained for arbitrary partial waves in 
\cite{kleb}. The greybody factor calculation for non-minimally
coupled fermions for arbitrary partial waves
agrees with the result found previously in
\cite{hosh}. Here, the matching of
the wavefunctions is non-trivial, as we
have to solve the wavefunctions in three
separate regions, near intermediate and
far to obtain the greybody factor. 
The vector gauge fields have not been
dealt with previously, and the emission rate
calculation is thus a prediction for the five
dimensional black hole. The near horizon
$SL(2,R)\times SL(2,R)$ symmetry imposes
some interesting restrictions on the 
one forms, which we exploit to obtain the
solutions of the vector equations of
motion.

Next we replace the entire near horizon geometry of
black string solution
by an effective  1+1-dimensional CFT which lies at a
finite distance from the horizon, i.e. at
$r\sim l\sim \sqrt{r_1r_5}$. Here $l$ is a measure of the size of the
near horizon geometry, and $r_1,r_5$ are
related to the charges of the black hole. 
A quantum mechanical 
calculation of the emission rate is done where a plane 
wave excites the operators of
the CFT. The correlators of the CFT operators 
are determined by the AdS/CFT correspondence
according to the prescription given in 
\cite{wit1,poly,mathur,muck,review,sfe,yi,wit2}. Unlike the other
calculations of the emission rates,
\cite{ohta1} where the $AdS$ bulk solution couples to the
operators, here it is the partial
wave components of the {\it plane wave}
which couple to the CFT operators and 
excite the CFT. The quantum mechanical calculation
using the correlators with their proper
normalisation constants reproduces the
emission rates {\it exactly}.

In the next section, to fix our notation we
review aspects of the six dimensional
compactification to $BTZ \times S^3$. 
We determine the
fermion and vector equations of motion
on the three dimensional black hole background,
and obtain the expression for the masses due to the orbital
angular momentum of the particles. 
In the third section, the equations
of motion for scalars, fermions and vectors 
are solved and the greybody factors
are determined. 
In the fourth section we determine the dimension
of the operators which couple to the above
particles and the corresponding correlators
with the exact normalisation. The emission
rates are then calculated by exciting the
operators using plane waves which are incident
on the CFT. The last section concludes with
a discussion.

\section{Five Dimensional Black Holes and Their Near Horizon Geometry}
\label{gbf}

The black hole solutions of string theory that we will consider arise
from the low energy effective action
of Type IIB string theory in $10$-dimensions,
by compactifying on $T^4 \times S^1$. 
The full $10$-dimensional metric is given by 
\cite{strom1,horo}
\begin{eqnarray}
ds^2~&=&~{f_1}^{-1/2}~
f_5^{-1/2}~
[ -dt^2 + (dx^5)^2 \nonumber \\
&+& \frac{r_0^2}{r^2}~\left(\cosh \sigma dt + \sinh \sigma dx^5 \right)^2 
+ f_1~dx_i dx^i ]  \label{10}\\
&+& f_1^{1/2}~ f_5^{1/2}~
\left[ \left( 1 - \frac{r_0^2}{r^2} \right)^{-1}
dr^2 + r^2 d\Omega_3^2 \right]~~,\nonumber  
\end{eqnarray}
where $x^5$ is along $S^1$ 
and $x^i~~,i=6,..,9$ are the coordinates on the $T^4$.
The functions $f_1$ and $f_5$ are given by:
$$f_1 = 1 + \frac{r_1^2}{r^2} ~~~,~~ f_5 = 1 + \frac{r_5^2}{r^2} ~~.$$

The resultant black hole metric in $5$-dimensions after Kaluza-Klein reduction
has six parameters, $r_1, r_5,r_0,\sigma,$~$V$ [volume of the $T^4$] and $R$
[length of the $S^1$].
In the case of the black hole obtained by wrapping $Q_5$
D-5 branes, $Q_1$ D-1 branes with
momenta $n$ along the 1-D brane 
the three charges of the black hole
viz. $Q_1$, $Q_5$, $n$ can be re-expressed as:
$$ r_1^2~=~\frac{g Q_1}{V}~,~~~r_5^2~=~gQ_5~,~~~\frac{r_0^2 \sinh 2\sigma}{2}
~=~\frac{g^2 n}{R^2 V}~.$$ 
The black hole horizon is at $r_0$. The non-zero field
strength in this background is given by:
\begin{equation}
H_{\mu\nu\rho}= \epsilon_{\mu\nu\rho}\frac{r~
r_1^2}{\left(r^2 + r_1^2\right)^2}\left(f_1
f_5\right)^{1/4};\;\;\;\; H_{abc}= \epsilon_{abc}\frac{r_5^2}{r^3}
\left(f_1f_5\right)^{-3/4} 
\label{field}
\end{equation}
Where $\mu,\nu..$ run over $t,x_5,r$ coordinates and $a,b,c$
denote the angular directions.
The metric (\ref{10}) 
has the interesting property that in the near horizon limit 
$r \rightarrow r_0$ and in the so-called dilute gas approximation $r_1,r_5 \gg
r_0, r_n$, it can be split up into three parts, 
\begin{equation}
ds^2~=~ds_{BTZ}^2  +  ds_{S^3}^2  + ds_{T^4}^2 
\label{decomp}
\end{equation}
where 
\begin{equation}
ds^2_{BTZ}~=~-\frac{\Delta^2}{ l^2 \rho^2} dt^2 + \frac{l^2\rho^2}{\Delta^2}
d\rho^2 + \rho^2 \left(d\phi - \frac{\rho_+ \rho_-}{l \rho^2} dt\right)^2
\label{btzm}
\end{equation}
$$\Delta^2~=~(\rho^2 -\rho_+^2)(\rho^2 - \rho_-^2)~~$$
is the metric of a $(2+1)$-dimensional $BTZ$ black hole, which is
a solution of Einstein equation in $3$-dimensions with a negative
cosmological constant $\Lambda = -1/l^2$ \cite{btz}. We have made a 
coordinate change $r^2= \rho^2 - \rho_-^2$ to get the above metric. 
The coordinate $\phi$, the parameter $l$ and the horizons 
of the BTZ black hole
$\rho_\pm$ are related to the $5$-dimensional black hole 
variables and parameters by the following relations:
\begin{equation}
\phi=x^5/l~~,~~\rho_+=r_0\cosh\sigma~~,\rho_-=r_0\sinh\sigma~~,~~
l^2=r_1 r_5~~.
\label{5dbtz}
\end{equation}
The part $ds^2_{T^4}$ is just the metric on the $4$-torus and
$ds_{S^3}^2$ is the metric on the three sphere with a constant 
radius $l$. The BTZ metric includes time, the
periodic $x^5$ direction and the
radial direction of the $5$-dimensional
black hole. 

The above decomposition forms the basis of the approach we are considering,
in which all thermodynamic properties of the 
black hole will be attributed to the 
`non-trivial' $BTZ$ part. Similar decompositions can be done in the 
case of black holes in other spacetime dimensions \cite{cvet,bal}.
The relevant near horizon part of the
metric thus preserves $SL(2,R) \times SL(2,R)$ symmetry which is
absent in the full five dimensional
geometry. In fact as shown in earlier
cases, the equation of motion of the
particles in the five dimensional black
hole background show this symmetry near
the horizon. The inclusion of the extra
direction $x^5$ does not affect this
property, since the extra dimension is a
Killing direction and does not change the symmetries of the
equations of motion.  

By compactifying the 10 D metric on $T^4$, the black
string solution in 6D is probed. This is a solution
of $N=8$ supergravity  in $6$-dimensions.  
In $D=6~,~N=8$ 
supergravity theory, the spectrum consists of 40 fermions
apart from 5 anti-self dual, anti-symmetric gauge fields, 16 vector fields, 
25 scalar fields and 8 Rarita-Schwinger fields. Out of these, as
seen in (\ref{field}), only one of the anti-symmetric gauge field strength
is non-vanishing apart from the metric background.
There is a $SO(5,5)$ global symmtery which gets broken due to
the black hole background to $SO(4)\times
SO(5)$. We look at certain particles in the spectrum,
namely the minimally coupled scalars and non-minimally coupled
fermions and gauge fields in 6D.
The scalars correspond to gravitons along
the $T^4$ direction. The scalars satisfy
ordinary Klein Gordon equation in 6D, and
on compactification on $AdS_3 \times
S^3$, are expanded as $\phi=
\sum \Phi(t,r,\phi)Y^{(L0)}$, where $Y^{(L0)}$ are
spherical harmonics on $S^3$. The 
equation of motion for scalar fields for the 
partial wave $L$ on $S^3$ satisfies the massive Klein-Gordon equation 
\begin{equation}
[\Box - M^2 ] \Phi = 0
\label{boxphi}
\end{equation}
in the $AdS_3$ spacetime with the mass $\mu$ given in terms of $L$ as
\cite{boer,sezgin}
$$ M^2 = \frac 1{l^2}L(L+2)$$

In the notation of \cite{tanii} the fermion equation
of motion is  
\begin{eqnarray}
\frac i2 \Gamma^M D_M \chi^{a \alpha} &-&\frac1{16} P_{\dot a 
~M}^{a}(\gamma^{\dot a})^{\alpha}_{ \beta}
\Gamma^M\Gamma^N\psi^{\beta}_{-N} 
-\frac{1}{24}F^{\dot a}_{MNP}\Gamma^{MNP}\gamma_{\dot
a}^{\beta \alpha}\chi^{a}_{\beta}  \nonumber \\
&+& F^{\alpha
\dot{\alpha}}_{MN}
\left[(\gamma^a)_{\dot\alpha\dot\beta}\Gamma^{MN}\Gamma^{P}\psi^{\dot \beta}_{+P} -
\frac14\Gamma^{MN}\chi_-^{\dot a \beta}(\gamma_a)^{\alpha}_{\beta}(\gamma_{\dot a})^{\dot
\alpha}_{\dot\beta}\right]=0 \label{tanii}
\end{eqnarray}
where $M, N..$ represent 6 dimensional world
index, $a$ and $\dot a$, $SO(5)\times
SO(5)$ vector index , $\alpha, \beta$
$SO(5)\times SO(5)$ spinor
index. The $+$ and $-$ sign denote the
chirality of the fermions. In the above
$P$ are related to the kinetic term of
the scalars, $F^a_{MNP}$ is related to
the three form field strength, $F^{\alpha
\dot \alpha}_{MN}$ are related to the
field strengths of the one form gauge
fields. Now we study the compactification of this
theory to $AdS_3 \times S^3$. From (\ref{field}) 
it can be seen that that the non-zero background fields, 
the three form
field strength is given in the near horizon limit by;
$H^a_{\mu\nu\rho}=(1/l)\epsilon_{\mu \nu \rho}\delta^{a5}$ 
and $H^a_{bcd}= (1/l)\epsilon_{bcd}\delta^{a5}$ 
where $\mu, \nu,...$ etc indicate the three
$AdS$ directions and $b,c,d...$ are the $S^3$ directions.
This gives the required equation $R_{\mu
\nu\rho\lambda}= -1/l^2(g_{\mu
\rho}g_{\nu\lambda}-
g_{\nu\rho}g_{\mu\lambda})$ for the $AdS_3$
directions and 
$R_{bcde}=1/l^2(g_{bd}g_{ce} - g_{cd}g_{be})$ for the $S^3$.
Next, we factorise the fermion field $\chi$ in terms of an 
undetermined function
of the $BTZ$ coordinates, times the harmonic functions on the
three-sphere. We also work in the
representation where 
$(\gamma^5)\chi=\chi$. In this linearised
approximation, the resultant expression is:
\begin{equation}
\Gamma^MD_M\chi -
\frac1{12}H_{MNP}\Gamma^{MNP}\chi=0
\label{fere}
\end{equation}
The expansion in harmonics of $S^3$ is of
the form $\chi = \sum \chi^{(p, \pm
1/2)}Y^{(p,\pm 1/2)}$, and obey
${\sb} Y^{(p,\pm1/2)}= \pm\imath (p
+1)Y^{(p,\pm1/2)}$, where $p$ is a half
integer, labeling the spin
representation. 
Plugging in this expansion in the equation of motion (\ref{tanii}),
and using the decomposition of $6$-dimensional $\Gamma^M$ 
matrices into $3$-dimensional ones as given in \cite{sezgin}, 
the two-component equation takes the form
\begin{equation}
\gamma^{\mu}D_{\mu}\chi'  + \frac{1}{l}\left(\mp (p
+1) - 1\right)\chi'=0
\end{equation}
Where $\chi^{(p,\pm1/2)}=\chi' $,
Which can be written as:
\begin{equation}
\gamma^\mu \left( \partial_\mu + \omega_\mu + 
\frac{1}{l}\left[ L + 1/2\right] \right)~\chi'~=~0~~~.
\label{fermq}
\end{equation}
where $p= L+ 1/2$, and we have chosen one of the eigenvalues
of the spherical harmonic 
(choosing the other sign gives $L + 1/2 +2$ for the mass term). 
The spin connections correspond to BTZ spacetime. Note that, here 
$L$ stands for the orbital angular
momentum, and in
\cite{adg}, the calculations were done
for $L=0$.  From the above it also follows that 
the lowest mass term in the $BTZ$ space time is non zero and
equals $1/2$. 
This is our basic set of equations for the determination fermionic
of the greybody factor. 
It is interesting that  on plugging the
three dimensional spin connections and
using the relations (\ref{5dbtz}), it can be
shown that this equation is the same as that of the fermionic fluctuations in
the background of the $5$-dimensional black hole in the near horizon limit
\cite{hosh}. We confine ourselves to
particles without any Kaluza-Klein momentum along the compact direction $x^5$.
In other words, the particles belong to
the $s$-wave sector with respect to the
$BTZ$ black hole. Inclusion of the
azimuthal quantum number along $x^5$ will imply charged
fermion emission in five dimensions.  

Similar decomposition can be made for the
vector equations of $D=6, N=8$
supergravity into $AdS_3 \times S^3$. The
exercise has been done in \cite{lars}.
Note that this vector couples to the
threeform, and hence its  linearised 
 equation of motion reduces to:
\begin{equation}
\nabla ^MF^{a}_{MN} - \frac16
\epsilon_N^{PQRST}(\gamma_a)^{\alpha}_{\beta}
F^{\beta}_{PQ}H^a_{RST}=0\\
\label{sixd}
\end{equation}
The gauge fields when expanded in the
spherical harmonics $A_{\mu}=
{\sum} A_{\mu}^{(L,\pm1)} Y^{(L,\pm1)}$ reduces
to:
\begin{equation}
\nabla^\nu \partial_{[\nu}A_{\lambda]} -
\frac1l \epsilon_\lambda^{~\nu\rho}\partial_{[\nu}A_{\rho]}
= \frac 1{l^2}L(L+2) A_\lambda
\label{veceom}
\end{equation}
wher we have dropped the indices
$(L,\pm1)$. These set of equations
correspond to a massive gauge field in
the BTZ background, and we solve for this
to get the required greybody factor.

\section{Greybody Factors}

In this section, we solve the scalar, fermion and vector equations of 
motion of the previous  sections to find the absorption cross-sections
of the black hole for these particles. 
Since we study particles of various spins,
a Newman-Penrose formalism would have been ideal
for the study of particle propagation on the
BTZ background. However, this has not been
developed in three dimensions, and we separately consider the various
equations of motion and find the solutions in the near horizon 
and in the asymptotic regions.

\subsection{Scalar Greybody Factor}
The scalar greybody factor for arbitrary
partial waves for the five
dimensional black hole was found in 
in \cite{kleb}. Here, we exploit the near-horizon ($BTZ$) 
geometry of the black holes to solve the scalar wave equations. 
As stated before, the massless scalar wave equation for
an arbitrary partial wave $L$
in the $5D$ background can be reduced to the massive Klein-Gordon 
equation in the BTZ background. 
This equation was solved for the massless case in \cite{sid}.

>From (\ref{boxphi}) and (\ref{btzm}), we we get the massive $s$-wave
scalar equation in $BTZ$ background: 
\begin{equation}
\frac{1}{\rho} \frac{d}{d\rho} \left( \frac{\Delta^2}{l^2 \rho}
\frac{d\Phi}{d\rho}\right) + \frac{\omega^2 l^2 \rho^2 }{\Delta^2} \Phi - M^2\Phi
=0~.
\end{equation}
Defining $$z = \frac{\rho^2-\rho_+^2}{\rho^2 -\rho_-^2}$$  and 
assuming $\Phi (x^\mu) \sim e^{i\omega t} \Phi (\rho)$, the equation takes the
form 
\begin{equation}
z(1-z)\frac{d^2\Phi}{dz^2} + (1-z) \frac{d\Phi}{dz} + \left[
\frac{A}{z} - B -\frac{M^2}{4(1-z)}\right]\Phi = 0
\end{equation}
where $$A=(\omega/4\pi T_H)^2, B=(\rho_-^2/\rho_+^2) A~~$$
and $$ T_H = \frac{\rho_+^2 - \rho_-^2}{2\pi l^2 \rho_+} $$
is the Hawking temperature of the BTZ black hole. 
Plugging in the ansatz
\begin{equation}
\Phi(z) = z^m (1-z)^n F[z]
\end{equation}
we get 
\begin{eqnarray}
z(1-z) \frac{d^2F}{dz^2} &+& \left[ (2m+1) - (2m+2n+1)z\right] \frac{dF}{dz}
 \nonumber   \\
&+& \left[ \frac{m^2+A}{z} + \frac{n(n-1) - M^2/4}{1-z} - (m+n)^2
-B\right] F =0 
\end{eqnarray}
Setting the coefficients of the $1/z$ and the $1/(1-z)$ terms to zero, as
required by the continuity with the solution very close to the horizon \cite{gbf},
the above equation reduced to the familiar hypergeometric equation
\begin{eqnarray}
z(1-z) \frac{d^2F}{dz^2} + [(2m+1) - (2m+2n+1)z] \frac{dF}{dz} -
[(m+n)^2 + B ] F =0~. 
\end{eqnarray}
Thus, the final solution is:
\begin{equation}
\Phi(z)= z^m (1-z)^n F[\alpha,\beta;\gamma;z]
\end{equation}
where
\begin{eqnarray}
m&=&-i\sqrt{A} ~~,~~
n=- \frac{L}{2} \nonumber \\
\alpha &=&-i(\sqrt{A} -\sqrt{B}) +n ~~,~~
\beta = -i (\sqrt{A} + \sqrt{B}) +n \\
\gamma&=& 1 - 2i\sqrt{A} \nonumber
\end{eqnarray}
and we have substituted $M^2 = L(L+2)/l^2$. 
The flux of particles into the black hole can be calculated from the formula
\begin{equation}
{\cal F}_0 = \frac{2\pi}{i} \left[ \frac{\Delta^2}{\rho}
\Phi^* \frac{d\Phi}{d\rho} - c.c. \right] 
\end{equation}

which yields

\begin{equation} 
{\cal F}_0 = 4\pi \omega l^2 \rho_+ 
\label{flux0} 
\end{equation}

Now to find the incoming flux at infinity, we have to solve the wave
equation at very large distances from the black hole, where space time is
almost flat. The corresponding wave equation is solved in the
six dimensional black string background, with the metric 
given in Eq.(\ref{10}), with $r \rightarrow \infty$. The solution is expanded as
$\sum \Phi(r)Y^{(L0)}$, where the $Y^{(L0)}$ are the spherical
harmonics on $S^3$. Using $\nabla^2 Y^{(L0)} = -L(L+2) Y^{(L0)}$, where 
$\nabla^2$ is the Laplacian on the $S^3$, the radial equation of motion follows:
\begin{equation}
\frac{1}{r^3}\frac{d}{d r}\left(r^3 \frac{d\Phi}{d r}\right) 
+\left[ \omega^2 - \frac{L(L+2)}{r^2} \right] \Phi = 0
\end{equation}
having the ingoing Bessel solution:
\begin{equation}
\Phi = \frac{1}{r}\left(AJ_{L+1} (\omega r) + B N_{L
+1}(\omega r)\right) 
\label{bessel}
\end{equation}

The asymptotic expansions of the Bessel functions yields the
following flux at infinity:

\begin{equation}
{\cal F}_\infty = 2\left[ |A|^2 + |B|^2 + i(A^*B - B^*A) \right]  
\label{finfty}
\end{equation}

Since the far solution should smoothly go over to the near horizon (BTZ)
solution, we investigate the nature of the solutions
near the region $r\sim l$, till which region we assume that the $AdS_3$
geometry is a good approximation to the black hole spacetime. From
Eq. (\ref{5dbtz}) and the dilute gas approximation, near $r \approx l
\gg r_0 \sinh\sigma$, we get $\rho \sim r$  and hence the angular parts of the
wavefunctions are the same. Thus, we simply
compare the radial wavefunctions. 
The intermediate region is obtained 
by setting $z \rightarrow 1$ and $r\omega<<1$ in the hypergeometric
and the Bessel solutions respectively to obtain the matching
condition \cite{abra,BM} :
\begin{equation} 
A = N^{-L/2} (L+1) (L!)^2 \left(\frac{2}{\omega} \right)^{L+1}~
\frac{\Gamma(\gamma)}{\Gamma(\gamma-\alpha)\Gamma(\gamma-\beta)}~~.
\label{A}
\end{equation}
where $N= \rho_+^2 - \rho_-^2 = r_0^2$. 
The other constant $B$ is much smaller by factor of $(N\omega^2)^{L}$,
and hence is neglected in the subsequent calculations.

A interesting point to note is that if we solve for the
scalar wavefunction in the asymptotic $AdS$ space, the solutions
are obtained as $\Phi= J_{L+1}(\omega l^2/\rho) +
N_{L+1}(\omega l^2/\rho)$, and thus 
at $\rho=r=l$, this has the exact
polynomial behaviour as the flat space wave functions as
the arguments of the Bessel functions both reduce to $\omega l$.
Although we do not use the scalar wave functions in asymptotically $AdS_3$
space to determine the greybody factor, it would be interesting to check
whether the above observation has a deeper significance, since 
the location $r=l$ has no apparent physical significance. 

The greybody factor is then evaluated using standard methods
of calculation of absorption crossections by taking the ratio
of the fluxes.  
Thus from (\ref{flux0}),  (\ref{finfty}) and ({\ref{A}}), the greybody factor
is:
\begin{eqnarray}
\sigma_{\mbox{abs}} &=& \frac{4\pi}{\omega^3} (L+1)^2 \frac{{\cal F}_0}{{\cal
F}_\infty} \nonumber \\
&=& \frac{2\pi}{(L!)^4} \left(\frac{\omega}{2}\right)^{2L}
\frac{N^{L+1}}{\omega} \sinh\frac{\omega}{2T_H}~ \label{semicl} 
\label{sgbf}\\
&\times& |\Gamma(1 + L/2 + i\omega/
 4\pi T_-) \Gamma(1+L/2 + i\omega/4\pi T_+)|^2 \nonumber
\end{eqnarray}

where $$\frac{1}{T_{-,+}} \equiv \frac{1}{T_H} \left(1- 
\frac{\rho_-}{\rho_+}\right)$$ and 
we have included the plane wave normalisation factor $\frac{4\pi}{\omega^3}
(L+1)^2$. We have also used the identity $|\Gamma(1-ix)|^2 =
\pi x/\sinh\pi x $. The above expression for the greybody factor reduces to 
the area of the black hole for $L=0~,~T_- \gg T_+~,$ and 
$\omega \rightarrow 0$ \cite{dgm}. 

\subsection{Fermion Greybody Factor}
We shall
solve the equation of
motion of the fermions equation (\ref{fermq}) on the BTZ
background, in a suitable set of
coordinates.
We define $\rho^2 = \rho_+^2 \cosh^2\mu - \rho_-^2 \sinh^2\mu$ 
and $ x^\pm = \pm \rho_\pm t/l \mp \rho_\mp \phi $
and assume the following form of the wavefunctions:
$$ \chi'_{1,2} = \frac{e^{i(k^+x^+ + k^-x^-)}}{ {\sqrt{\cosh\mu \sinh\mu}}}~\psi_{1,2},~$$ 
where $(1,2)$ refer to the two components of the spinor. 
The spin connections for the $BTZ$-metric are:
$$  \omega_{x^+} = -\frac{1}{2l} \cosh \mu \sigma^{01} ~~~,~~
\omega_{x^-} = \frac{1}{2l} \sinh \mu \sigma^{21} $$
The equation of motion for $\psi$ takes the
following form:
\begin{equation}
\gamma^1\partial_{\mu}\psi + \gamma^0\frac{ilk^+}{\sinh{\mu}}\psi +
\gamma^2\frac{ilk^-}{\cosh\mu}\psi + \left( L +\frac12 \right)\psi=0~.
\end{equation}
Here we work in the representation, $\gamma^1=\sigma^1, \gamma^0=\imath \sigma^2,
\gamma^2=\sigma^3$.
Then we define a new set of wavefunctions $\psi_{1,2}'$ as
\begin{eqnarray}
\psi_1 + \psi_2 ~&=&~ \left( 1 - \tanh^2\mu\right)^{-1/4}{\sqrt{1+\tanh\mu}}~
\left(\psi_1' + \psi_2'\right)\label{eqn1} \\
\psi_1 - \psi_2 ~&=&~ \left( 1 - \tanh^2\mu\right)^{-1/4}{\sqrt{1-\tanh\mu}}~
\left(\psi_1' - \psi_2'\right)~~
\label{eqn}
\end{eqnarray}
whence the Dirac equation assumes the form:
\begin{eqnarray}
(1-y^2) d_y \psi_2' - i\left(\frac{k^+}{y} + k^-y\right)~\psi_2' &=&
-[L+1  + i(k^+ + k^-)] \psi_1'  \\
(1-y^2) d_y \psi_1' + i\left( \frac{k^+}{y} + k^- y\right)~\psi_1' &=& 
-[L+1 - i(k^+ + k^-) ] \psi_2' 
\end{eqnarray}
where we have defined $y =\tanh\mu$. 
Next, we choose the following ansatz
\begin{equation}
\psi_{1,2}'~=~B_{1,2} z^{m_{1,2}} (1-z)^{n_{1,2}} F_{1,2} (z)~~~,
\end{equation}
where $B_{1,2}$ are arbitrary constants $z=y^2$ and $F(z)$ are yet undetermined 
solutions. Substituting in the Dirac equations, separating the equations
for $\psi_1'$ and $\psi_2'$ and demanding continuity of this solution with the solution
obtained very close to the horizon, we finally obtain the following 
hypergeometric differential equations for $F_1(z)$ and $F_2(z)$ :

\begin{eqnarray}
z(1-z) \frac{d^2F_i}{dz^2} &+& [ (2m_i+1/2) - (2m_i + 2n_i +3/2)z]\frac{dF_i}{dz} \nonumber \\
&-&[m_i(m_i+1/2) + n_i(n_i+1/2) + 2m_in_i - \frac{il k_- - l^2k_-^2}{4} ] F_i =0 
\end{eqnarray}
The constants $m_i,n_i$, the hypergeometric parameters $\alpha_i,\beta_i,\gamma_i$ 
and the integration constants $B_i$ are tabulated below

\begin{eqnarray}
m_1~&=&~\frac{1+ il k^+}{2} = m_2 + 1/2 \nonumber  \\
n_1~&=&~-\frac{1}{2}(L+1)~=~n_2 \nonumber \\
\alpha_1~&=&~ m_1 + n_1 + \frac{1}{2} + \frac{ il k_-}{2} ~=~ \alpha_2 +1 \nonumber \\
\beta_1~&=&~ m_1 + n_1 - \frac{il k^-}{2}~=~\beta_2 \\
\gamma_1~&=&~2 m_1 + \frac{1}{2}~=~\gamma_2 +1 \nonumber \\
B_2~&=&~-\left[ \frac{\gamma-1}{\alpha - (2n+1)}\right] B_1 \nonumber
\end{eqnarray}
In our subsequent calculations, we shall
normalise $B_1=1$. This solution is an
exact solution for the BTZ space time and it approximates the fermionic wave function near the
horizon.

The flux into the black hole can be calculated using the $\rho\rightarrow \rho_+, z\rightarrow
0$  limit of this solution. The flux of particles entering the horizon is 
\begin{equation}
{\cal F}_0 = {\sqrt {-g}} J^\rho |_{\rho_+} = 
{\sqrt{-g}} {\bar \psi}e_1^\rho \gamma^1 \psi~~.
\end{equation}
Substituting the above solutions it is clear that $\psi_2'$ dominates the flux, 
and the 
latter turns out to be
\begin{equation}
{\cal F}_0 = N \left| \frac{\gamma-1}{\alpha -(2n+1)}\right|^2~.
\end{equation}

Now, to find the incoming flux at infinity, we solve the radial 
Dirac equation in the $6$-dimensional metric (\ref{10}) away from the horizon, 
i.e. taking $r_n^2/r^2, r_0^2/r^2 \rightarrow 0$. Then the metric
assumes the following form:
\begin{equation}
ds^2= -\frac{1}{\sqrt{f_1f_5}} dt^2 +
\frac{1}{\sqrt {f_1f_5}} dx_5^2 +
\sqrt{f_1f_5}\left(dr^2 + r^2 d\Omega^2\right)
\end{equation}
The spin connections for this metric are:
$$w_{t}^{01}= -w_{x_5}^{51}=
\frac{1}{4\sqrt{f_1f_5}}\left[\frac{r_1^2}{r^3
f_1} + \frac{r_5^2}{r^3 f_5}\right],
\;\;w_b^{i1}= \frac12 - \frac{r}{4}\left[\frac{r_1^2}{r^3f_1} + \frac{r_5^2}{r^3f_5}\right]$$
where $b$ stands for the $S^3$ world indices and
$i$ the $S^3$ tangent space index.
Now, in six dimensions the
wavefunction is a four component chiral spinor. We
start with the appropriate equation of motion in 6D
as given in 
(\ref{fere}). Including all the terms, the equation of
motion is:
\begin{equation}
\left[ \left(f_1f_5\right)^{1/2}\Gamma^0\partial_0 + \Gamma^1\left(\partial_r +
\frac{3}{2r} + \frac{1}{8}d_r (\ln (f_1f_5))\right) +
\left(f_1f_5\right)^{1/2}\Gamma^{5}\partial_{x_5} + 
\Gamma^{b}D_b\right]\chi + g(r)\chi=0 
\label{f4}
\end{equation}
Where $D_b= d_b + w_b$, where $b$ denotes the
$S^3$ directions, and $w_b^{ij}\sigma_{ij}$ is the spin
connection with the $i,j$ indices running over
tangent space $S^3$ indices only. The function
$$g(r)= \frac1{12} \Gamma^{MNP}H_{MNP}=  -\frac{1}4\left[ d\ln
(f_1f_5)\right]\sqrt{\frac{f_1}{f_5}}$$
Using the decomposition of $\Gamma$ matrices
into SO(2,1) and SO(3) parts we can separate out
the equation of the components of the 6D chiral
wavefunction into two sets of two component wave
functions \cite{sezgin}. 
Again we expand in terms of the spherical harmonics on $S^3$ as:
$\chi= \sum \chi'(x_{\mu}) Y$, where $\chi'$ are two
component wave functions. Further,
$\chi'=  e^{i(\omega t - m\phi)}~
\left(f_1f_5\right)^{-1/8}r^{-3/2}\psi''(r) $ is defined. 
Then from (\ref{f4}), we get:
\begin{equation}
\left[(f_1f_5)^{1/2}\gamma^0\partial_t + \gamma^1\partial_r +
(f_1f_5)^{1/2}\gamma^{2}\partial_{x_5}\right]\psi'' =  \left[\frac{(L +
3/2)}{r} - g(r)\right]\psi''
\end{equation}
Separating the components gives us the 
equations:
\begin{eqnarray}
\left(d_r - (f_1f_5)^{1/2}\imath \omega\right)\psi''_1 &= &
\left(-\frac{L +3/2}{r} +g(r) - (f_1f_5)^{1/2}\imath
m\right)\psi''_2\\
\left(d_r + (f_1f_5)^{1/2}\imath \omega\right)\psi''_2 &= &
\left(-\frac{L +3/2}{r} +g(r) + (f_1f_5)^{1/2}\imath
m\right)\psi''_1
\end{eqnarray}
Defining,
$\psi''_1 + \psi''_2 = (f_1f_5)^{-1/4}\psi^{+}$ and 
$\psi''_1 - \psi''_2=(f_1f_5)^{1/4}\psi^-$ 
and with the additional approximation
$g(r)= - 1/4 d_r\ln(f_1f_5)$, the equations reduce to:
\begin{eqnarray}
\left(d_r + \frac{L+3/2}{r}  \right)~\psi^+ &= &i(f_1f_5)\omega \psi^-
\label{couple}\\
\left(d_r - \frac{L+3/2}{r}\right )\psi^- &= &i\omega \psi^+
\end{eqnarray}
where we have put $m=0$. 
The second order differential equation has the following form
for $\psi_-$
\begin{equation}
\left[d^2_r - \frac{(L+3/2)(L + 1/2)}{r^2} + \omega^2(f_1f_5)\right]
\psi^-=0
\end{equation}
We solve this equation in two regions: $r \sim l$ and $r \geq
l$.

\noindent
{\bf Intermediate Region}\\
In the first region $r \sim l$, which we call the intermediate region, 
we take $ \omega^2f_1f_5 \approx \omega^2(r_1^2 + r_5^2)/r^2 + \omega^2l^4/r^4$ 
for low energy emissions. The differential equation in
terms of $x=1/r$ has the form:
\begin{equation}
\left[d^2_x + \frac{2}{x}d_x - \frac{(L +3/2)(L +1/2) - (r_1^2
+ r_5^2)\omega^2}{x^2}
+ \omega^2l^4\right]\psi^-=0
\end{equation}
The solution for the above differential equation
is the Bessel function $x^{-1/2}Z_{\nu}(\omega
l^2x)$ where, $\nu= \sqrt{(L +1)^2 -
(r_1^2 + r_5^2)\omega^2}\approx L+1$ for low energy
emissions $\omega l\ll 1$. Hence
explicitly the solutions are:
\begin{equation}
\psi^-= \sqrt{r}\left[a_1 J_{L +1}\left(\omega l^2/r\right)
+ a_2 N_{L+1}\left(\omega l^2/r\right)\right] ~.
\end{equation}
And the coupled
differential equation for $\psi^+$ yields:
$$\psi^+= \frac{il^2} {r^{3/2}}\left[a_1 J_L \left(\omega
l^2/r\right) + a_2 N_L\left(\omega l^2/r\right)\right]$$
For $r< l$, the function $f\approx l^4/r^4$ and
in the limit we are considering, i.e.
$\omega l\ll 1, r\sim l$, we can do a
small argument expansion of the bessel
function. Hence
\begin{eqnarray}
\psi''_1 + \psi''_2 &\approx &\frac{i l}{\sqrt r}\left[a_1
\frac1{L!}\left(\frac{\omega l^2}{2r}\right)^{L} + a_2
(L-1)!\left(\frac{2r}{\omega l^2}\right)^{L}\right]\\
 \psi''_1 -
\psi''_2 &\approx&\frac{l}{\sqrt r}\left[ a_1
\frac{1}{(L+1)}!\left(\frac{\omega l^2}{2r}\right)^{L+1} + a_2L!\left(\frac{2r}{\omega
 l^2}\right)^{L+1}\right].
\end{eqnarray}
Which gives the leading order behavior of
\begin{equation}
\chi'_{1(2)} \sim a_2 L!\left(\frac{\omega
l^2}{2}\right) r^{L-1/2}$$
\label{in}
\end{equation}
For $r> l$, $f_1f_5\approx 1$ and
hence
 the above wavefunctions go to:
\begin{eqnarray}
\psi''_1 + \psi''_2 &\approx &\frac1{r^{3/2}}\left[a_1
\frac1{L!}\left(\frac{\omega l^2}{2r}\right)^{L} + a_2
(L-1)!\left(\frac{2r}{\omega l^2}\right)^{L}\right]\\
 \psi''_1 -
\psi''_2 &\approx &\sqrt{r}\left[ a_1
\frac1{(L+1)!}\left(\frac{\omega l^2}{2r}\right)^{L+1} + a_2L!\left(\frac{2r}{\omega
  l^2}\right)^{L+1}\right].
\end{eqnarray}
Which gives the wavefunction in the
leading powers of $r$ as:
\begin{equation}
\chi'_{1(2)}= a_2 L!\left(\frac{\omega
l^2}{2}\right)^{L+1} r^L
\label{if}
\end{equation}

\noindent
{\bf Far region:} \\
For 
$r>r_1,r_5$, we approximate $f_1f_5\approx1$, 
and the second order differential equation for $\psi^+$ is:
\begin{equation}
d^2_r \psi'^+ + \left[\omega^2 + \frac{(L+2)^2 -1/4}{r^2}\right]\psi'^+=0
\end{equation}
This has the solution: 
\begin{equation}
\psi'^+ = \sqrt
{\omega r}\left(a'_1 J_{L+2}(\omega r) + a'_2
N_{L+2}(\omega r)\right)
\end{equation}
Now, we can use this solution in the coupled equation (\ref{couple})
and get
$$\psi^-= \imath \sqrt{\omega r} \left[ a'_1 J_{L+1} + a'_2
N_{L+1}\right].$$ 
We now see, how the wave functions behave and obtain matching
conditions for their smooth joining.
Using the expansion for Bessel functions we obtain
the leading order behavior of the wavefunctions as:
$r\sim l$:
\begin{equation}
\chi'_{1(2)}  = a'_1\frac{\sqrt{\omega}}{(L+1)!}\left(\frac{\omega}{2}\right)^{
L+1} r^{L} 
\label{fi}
\end{equation}
and For $r\rightarrow \infty$ the
asymptotic expansion of the bessel
functions become important and the 
wavefunctions go as;
\begin{equation}
\chi'_{1(2)}= a'_1\frac 1{\sqrt{2\pi
r^3}}e^{-\imath \omega r}
\label{far}
\end{equation}
The flux at infinity entering the black hole spacetime is calculated from the asymptotic
expansions of the Bessel functions, which is given by
\begin{equation}
{\cal {F}}_\infty = \frac{|a'_1|^2}{2\pi}
\end{equation}

\noindent
{\bf Matching:}\\
To compare with the near horizon wave function solved in the
$x^+,r, x^-$ coordinates, we have to use the properties of
the spinor under such transformations from $t, r,\phi$
coordinates. This gives a rotation on the two component wavefunction by a matrix
:
$$[\cosh\left({\frac{\xi}2}\right) + i\sinh\left({\frac{\xi}2}\right)\sigma_2]\chi,\;\;\; 
\cosh{\xi/2}= \sqrt{\rho_+ + \rho_-}/N^{1/4} + \sqrt{\rho_+ -
\rho_-}/N^{1/4}~.$$   
The near horizon solution, when extrapolated to $z \rightarrow 1$ 
(keeping the leading term in the expansion) is
\begin{equation}
\chi'_{1(2)} \rightarrow \sqrt {\rho_+ - \rho_-}L! 2^{1/2}N^{-L/2} G ~\rho^{L-1/2}~,
\label{ni}
\end{equation}
where 
$$G = \frac{\Gamma(3/2 + i\omega/2\pi T_H)}{\Gamma([L+3]/2 + i\omega/4\pi T_+) \Gamma([L+2]/2 -
i\omega/4\pi T_-)} ~,$$
Thus comparing with the intermediate
solutions and then with the far solution
using equations (\ref{in},\ref{if},\ref{fi}) we get:
\begin{equation}
a'_1= 2^{L +3/2}(L+1)L!^2 \omega^{-L-3/2} N^{-(L/2)}  G ~~.
\end{equation}

Substituting in $F_\infty$ we finally get 
\begin{eqnarray}
\sigma_{\mbox{abs}} &=&
\frac{\pi (L+1)(L+2)}{\omega^3}~\frac{F_0}{F_\infty}\nonumber
\\
 &= &\frac{\pi 
(L+2) N^{L+1}}{2(L+1)(L!)^4 \left(\rho_+ - \rho_-\right)} \left(\frac{\omega}{2}\right)^{2L}
 \nonumber \\
&\times& \cosh
(\omega/2 T_H)\left|\Gamma( L/2 + 1/2 +i\omega/4\pi T_+)\Gamma(L/2 +1
+i\omega/4\pi T_-)\right|^2 \label{fermigbf}
\label{fgbf}
\end{eqnarray}
where, we have used the fact that $|\Gamma(1/2 +ix)|^2=
\pi/\cosh\pi x$
and we have multiplied by the appropriate plane wave
normalisation \cite{hosh}. The
wavefunction corresponding to the $S^3$
spinor $Y^{p,-1/2}$ gives rise to a
greybody factor with $T_+\rightarrow T_-$
and vice-versa. Hence the total greybody
factor is a sum of two terms, one due to
each set of two component fermions.

\subsection{Vector Greybody Factor}

The vector equation of motion is given in (\ref{veceom}). The
higher partial wave in five dimensions gives a mass term for
the gauge field in three dimensions. In
addition, there is another set of equations as
given in \cite{lars} :
\begin{equation}
\epsilon_\lambda^{~\nu\rho} \partial_\nu A_\rho = -\frac{L}l A_\lambda
\label{eps}
\end{equation}
This is derived from the representation theory of one forms on SL(2,R)
manifolds. Since the BTZ space is locally anti-de Sitter,
whose covering group is SL(2,R)$\times$ SL(2,R), the equation
of motion gets supplemented by the above. On substituting the above equation 
in (\ref{veceom}), the vector equation of motion reduces to (for
convenience, we
set the AdS radius $l=1$ in the rest of this section) :
\begin{equation}
\nabla^\nu \partial_{[\nu}A_{\lambda]} 
= L^2 A_\lambda
\label{eom}
\end{equation}
It is to be observed that (\ref{eom}), can now be derived from
(\ref{eps}) by operating with $\nabla$ on both
sides. There is also the consistency condition :
\begin{equation}
A^\nu_{;\nu}~=~0~.
\label{gauge}
\end{equation}
We would like to solve the above equations of motion in the background of the BTZ black
hole. In the coordinate system $(\mu, x^+, x^-)$ that we had adopted previously,
the $+$ and $-$ components of (\ref{veceom}) can be written as:
\begin{eqnarray}
\partial^2 A_+ + (\tanh\mu -\coth\mu) \partial_\mu A_+ + 2\coth\mu \partial_+ A_\mu -
2\tanh\mu \partial_{[-}A_{\mu]} &=&L(L+2) A_+ \label{b+} \\
\partial^2 A_- - (\tanh\mu -\coth\mu) \partial_\mu A_- + 2\tanh\mu \partial_- A_\mu -
2\coth\mu \partial_{[+}A_{\mu]} &=&L(L+2) A_- \label{b-} 
\end{eqnarray}
where $\partial^2 \equiv g^{\alpha\beta} \partial_\alpha \partial_\beta 
=\partial^\mu
\partial_\mu + \partial^+\partial_+
+\partial^-\partial_-$, we have taken $\epsilon^{+ \mu  -} =1$ and have used the
gauge condition (\ref{gauge}). 
Defining
\begin{equation} 
A_{1,2} = A_+ \pm A_-
\label{def}
\end{equation}
it is clear that the equation for $A_2$ gets decoupled by 
adding (\ref{b-}) and (\ref{b+}). To decouple the equation for $A_1$, 
we use the equations
(\ref{eps}) to substitute for the $A_\mu$ terms in (\ref{b+}) and (\ref{b-}). 
As a result, we get the following equations for the $A_1$ and $A_2$ (These set of
equations can also be derived directly from (\ref{eom}):
\begin{equation}
\partial^2 A_i + (\tanh\mu + \coth\mu) \partial_\mu A_i = (L^2 - 2\epsilon_iL)A_1~~,
\label{i}
\end{equation}
where $i=1,2~~,~~\epsilon_1= - 1~~,~~\epsilon_2=1$. Next, we substitute the solution
$$A_i=e^{ik_+x^+ + k_-x^-}A_i(\mu)~~,$$
which is consistent with the isometries of the metric. Substituting in
(\ref{i}), and defining $z \equiv \tanh^2\mu$ we get:
\begin{equation}
z(1-z) \frac{d^2A_i}{dz^2} + (1-z) \frac{dA_i}{dz} + \left[ \frac{k_+^2}{4z} -
\frac{k_-^2}{4} -\frac{L^2 - 2\epsilon_iL}{4(1-z)} \right]~A_i~=0~.
\label{hg1}
\end{equation}
Next, we substitute the ansatz 
$$A_i~=~e_i z^{m_i} (1-z)^{n_i} F_i(z)$$
in the above equation to obtain ($e_i$ s are constants)
\begin{eqnarray}
z(1-z) \frac{d^2F_i}{dz^2} + [(1+2m) - z(1+2m+2n)]~\frac{dF_i}{dz} \nonumber \\
+ \left[\frac{m^2+k_+^2/4}{z} + \frac{n(n-1) - (L^2 - 2\epsilon_i L)/4}{1-z}\right]~F
 \\
-\left[(m+n)^2 + \frac{k_-^2}{4} \right]~F~=0~ 
\label{hg2}
\end{eqnarray}
Continuity with the corresponding wave equations very close to the horizon 
$( z \rightarrow 0)$ gives harmonic
solutions in $\log z$ of the form $A_i =
e_i^{in} e^{\imath k_1 logz} + e_i^{out} e^{-\imath
k_2 \log z}$. To obtain a ingoing
solution, we put $e_i^{out}=0$. To ensure
that (\ref{hg2}) smoothly joins with
this, we determine $m$ and $n$ and find
that the coefficients of $1/z$ and
$1/(1-z)$ terms vanish. 
The residual part of
(\ref{hg2}) is simply the hypergeometric differential 
equation. Thus the functions $F_i(z)$ are
the hypergeometric functions $F[a_i,b_i;c_i;z]$ and the complete solution
for the gauge potentials can be written as

\begin{equation}
A_i~=~e_i z^{m_i} (1-z)^{n_i} F[a_i,b_i;c_i;z]~~.
\label{hg3}
\end{equation}

We can express the various parameters in terms of $k_\pm$ and $L$ :
\begin{eqnarray}
m_i&~=~& -i\frac{k_+}{2}  \nonumber \\
n_1~&=&~\frac{L}{2} + 1~~~.~~~n_2~=~\frac{L}{2}  \nonumber \\
a_1&=&~ -\frac{i}{2}(k_+ - k_-) +\frac{L}{2} + 1~~~,
~~~b_1~=~-\frac{i}{2}(k_++k_-) +\frac{L}{2}  + 1 \label{a1} \\
a_2&=&a= -\frac{i}{2}(k_+ - k_-) +\frac{L}{2}~~~,
~~~b_2 =~ b= -\frac{i}{2}(k_++k_-) +\frac{L}{2}    \nonumber  \\
c_i&=c=&~1+2m_i  \nonumber
\end{eqnarray}
$A_\pm$ can now be determined from the definitions (\ref{def}) and the
solution for $A_\mu$ can be constructed from the $\mu$-component of (\ref{eps}):
\begin{equation}
A_\mu~=~\frac{1}{L\cosh\mu\sinh\mu}\partial_{[+}A_{-]}
\label{mu}
\end{equation}
The important point to note is that the
two components $A_i$ satisfy equations
which are scalar equations in the BTZ
background. The spin dependence of the
solutions is not obvious. 
The constants $e_1$ and $e_2$ are not independent by virtue of the
auxiliary equations (\ref{eps}) and
the consistency conditions.
To determine the ratio, 
we use the equation with $\mu= +$
in (\ref{eps}).   
\begin{equation}
-\tanh\mu \left(\partial_{\mu} A_- -
\partial_{-}A_{\mu}\right) = L A_+
\end{equation}
On substituting $A_{\mu}$ from
(\ref{eps}), and going the $z$
coordinates, the equation reduces to
in terms of $A_1$ and $ A_2$ as,
\begin{equation}
\left[ 2 z d_z - \frac{2k_-k_2}{L} +
\frac{L}{1-z}\right] A_1= \left[ 2 z d_z
- \frac{2k_- k_2}{L} -
\frac{L}{1-z}\right]A_2
\end{equation}
On substituting the solutions for $A_i$, the above
simplifies to:
\begin{eqnarray}
e_1\left[ \frac{2 abz}{c} F(a+1,
b+1;c+1;z) + \left( a+b -
\frac{2k_-k_2}{L}\right)
F(a,b;c;z)\right]&& \nonumber\\
= e_2(1-z)\left[ \frac{2z(a+1)(b+1)}{c}
F(a+2,b+2;c+1;z) + \left( a+b+2 -
\frac{2k_-k_1}{L}\right.\right.&& \nonumber\\
 - \left.\left.\frac{2(L +
1)}{(1-z)}\right) F(a+1,b+1;c,z)\right]&&
\end{eqnarray}
Using a series of recursion relations, 
we get some simplified expressions \cite{abra}. 
The final expression is written below:
\begin{eqnarray}
e_1\left[ 2b F(a,b+1;c;z) + \left( a-b -
\frac {2k_-k_2}{L}\right)F(a,b;c;z)\right]&& \nonumber\\
= \frac{e_2}{a}\left[ 2(b-L)a + (a-b)L +
2k_-k_1\right]F(a,b+1;c;z) &&\\+
\frac{e_2}{a}
\left( a -b +
2k_-k_1/L\right)(a-L)
F(a,b,c;z)&&\nonumber\\
\end{eqnarray} From the above, the ratio of constants
are now easily determined to be:
\begin{equation}
\frac{e_2}{e_1}= -\frac{b^*}{a}
\end{equation}
where $k_{1,2} \equiv [k_+ \pm k_-]/2$. 
Plugging in this ratio of constants into the solutions and using
appropriate recursion relations, the wavefunctions can be written as
\begin{eqnarray}
A_+&= &\frac{e_2}{2b^*}(1-z)^{L/2}z^{ilk^+/2}\left[ -L
F(a,b+1;c;z) + (L - ik^+)F(a,b;c;z)\right]\\
A_-&= &-\frac{e_2}{2b^*}(1-z)^{L/2}z^{ilk^+/2}\left[ L
F(a,b+1;c;z) + ik^-F(a,b;c;z)\right]
\end{eqnarray}

In the above, the solution is actually
the real part of the wave function
determined above. 
The flux of the vector field at the horizon of the black hole is calculated 
using the energy momentum tensor for the
massive vector field. Since our
wave function is Re $A_i$, the energy
momentum tensor which involves products
of the fields will have the square terms
proportional to $e^{2i\omega t}$ and $e^{-2\imath
\omega t}$. Under time averaging, these
terms go to zero, and hence the steady
rate of particle influx is given by cross terms
:
\begin{equation}
T_{\nu\lambda}~=~-\frac{1}{4}(|F|^2 + 2m^2|A|^2) g_{\nu\lambda} +
F_{\nu\sigma}F^{*~\sigma}_{\lambda} +m^2 A_\nu A^*_\lambda 
\label{flux}
\end{equation}
Where $m$ stands for the mass. For our purposes $m^2=L(L+2)$.
To determine the flux, we incorporate the red-shift factor  
and integrate over the
horizon area to get: 

\begin{equation}
{\cal{F}}_0~=~ \frac{l^2 N^2 L}{2\rho_+}
k_+^2\left|\frac{e_2}{b}\right|^2\Omega
\label{f0}
\end{equation}
where $N \equiv \rho_+^2 - \rho_-^2, k^{+}= \omega/(2\pi l T_H)$, and we have restored the
radius of anti-desitter space. Also $\Omega = 8\pi^2$ denotes the
factors which come from the angular integrals. 
  Note that the flux vanishes for $L=0$, since the
latter is a not a dynamical mode \cite{lars}.

Before determining the waveform at infinity, we solve (\ref{eom})
in the asymptotic $AdS_3$ metric in the coordinates
$ (t,\rho,\phi)$as an interesting exercise, as it sheds light on the boundary behavior
of the wavefunction in the BTZ geometry. The wavefunctions, $A_i= e^{\imath
\omega t}B_i$ are solved, with the help of (\ref{eps})
as:
\begin{equation}
B_i(x) ~=~ {\sqrt{\rho}} \left[ c_i J_{\nu_i}(\frac{\omega
l^2}{\rho}) + d_i N_{\nu_i} ( \frac{\omega l^2}{\rho}
) \right]
\label{bessel1}
\end{equation}

where $J_\nu$ and
$N_\nu$ are Bessel
functions of the first and second kind
respectively, $~\nu_1=L-1~,~\nu_2=L+1$ and $c_i,d_i$
are arbitrary constants. Further, consistency with the equations (\ref{eps})
requires
that $c_1=-c_2 \equiv c$ and $d_1=-d_2\equiv d$.

To determine the wavefunction at asymptotic infinity which
joins with the BTZ wavefunction, we need to look at the
vector equation of motion in six dimensions. As given in 
Eq. (\ref{sixd}), the vector equation of motion
involves all the other $A_a$ components which are
scalar in the $t,r,\phi$ plane as well as the
$H_{MNP}$ three form field strength in six dimensions.
Since we are interested in that part of GBF which is due to the three dimensional vectors, we 
take $N=\mu$ in eqn \ref{sixd} and take the limit
$r\rightarrow\infty$. The equation of motion reduces to:
\begin{equation}
\nabla^{\nu} F_{\mu \nu} + \nabla^{a}F_{a \mu}=0
\end{equation}
Where we have kept terms of O$(1/r^2)$.
 In six dimensions $H_{MNP}= \epsilon_{ijkl}d_lf_5$ where
$\epsilon_{ijkl}$ is the flat space epsilon tensor along the
four non-compact directions $x_i$, which gives the second
term in equation (\ref{sixd})
to be order $(1/r^3)$ form (\ref{field}) and hence
can be ignored. In the gauge $\nabla^M A_M=0$, we
assume that $\nabla^{\nu} A_{\nu}= \nabla^a
A_a=0$. The main observation is that the $A_a$'s
decouple in this gauge. For the wavefunctions
$A_{\mu}= e^{i\omega t}e^{im\phi}A'_{\mu}(r)/r^{3/2}$, the
equation of motion for the $m=0$ case is of the form:
\begin{equation}
\partial^2_r A'_{t,\phi} + \left[ \omega^2 -
\frac{ (L+1)^2 -1/4}{r^2}\right] A'_{t,\phi}=0
\end{equation}
The solutions are:
\begin{eqnarray}
A'_t& = &\frac1r\left[a_1 J_{L+1}(\omega r) + a_2N_{L+1}(\omega
r)\right]\\
A'_{\phi}& = &\frac1r\left[a_1' J_{L+1}(\omega r) + a_2'
N_{L+1}(\omega r)\right]\\
A'_r & = &\frac1{r^3} (-\imath \omega)\int_r^{\infty}
r'^3 A_t(r')dr'
\label{bessel2}
\end{eqnarray}
It is interesting to note that the wavefunctions
determined here do not share the exact polynomial nature
of the wavefunction obtained in (\ref{bessel1}) at $r=\rho=l$,
as in the case of scalars. The reason behind this is that due to the loss
of $SL(2,R)\times SL(2,R)$ symmetry, the equations (\ref{eps}) are no longer valid for the
asymptotic metric. 
Thus the wavefunctions match with each other only in leading order in 
$\omega r$. 
Let us find the relation
between the coefficients of the solutions (\ref{hg3}) and (\ref{bessel2}) for which, 
we compare the two solutions in the region $z\rightarrow 1$, and $r\omega <<1$.
Using standard results for the behaviour of hypergeometric functions as $z\rightarrow
1$, we find the leading behaviour of the wave functions as
\cite{BM}: 
\begin{eqnarray}
A_+,A_- \rightarrow   \frac{e_2}{2 b^*}\frac{(N)^{-L/2}L\Gamma(L)
\Gamma(c)}{\Gamma(a)\Gamma(b +1)}\rho^{L} 
\end{eqnarray}
We match the solutions with the far region wavefunctions using
the relation: $A_t l = \rho_+ A_+ - \rho_-A_-$ which gives:
\begin{equation}
a_1 l= a_1'= \frac{e_2}{2 b*}\left(\rho_+ - \rho_-\right)N^{-L/2}\left(\frac{\omega}{2}\right)^{-(L
+1)}\Gamma(L +2)\Gamma(L + 1) E_1
\end{equation}
Where $E_1= \Gamma(c)/(\Gamma(a)\Gamma(b +1))$.
The other constants are negligible and hence ignored. The
solutions go as $A'_i\sim \sqrt{1/ 2\pi\omega} e^{-i \omega r}$ at
large distances. The flux, determined from equation (\ref{flux}) is: 

\begin{equation}
{\cal {F}}_{\infty}~=~-\frac{\omega}{2\pi} l^2|a_1|^2~ \Omega.
\label{f1}
\end{equation}
Taking the ratio of the near horizon and asymptotic fluxes 
(\ref{f0}) and (\ref{f1}) and
using the above relations for the ratio of the constants, 
we finally get the probability of absorption of the $L^{th}$ partial wave as
\begin{equation}
{\cal {P}}_L = \frac{{\cal {F}}_0}{{\cal {F}}_\infty}= \frac{\pi L k^{+2} N^L \omega^{2L
+1}}{ l^2\rho_+ 2^{2L} (\rho_+ -\rho_-)^2 (\Gamma(L+2)
\Gamma(L+1))^2 |E_1|^2} 
\end{equation}
This is the general result for the partial wave $L$. It is clear that the evaluation of
the gamma-functions will give rise to the familiar form of the greybody factor with
thermal distribution functions corresponding to two incoming particles and one outgoing
particle. The latter always is always associated with a Bose distribution function, as
can be seen from the relation $|\Gamma(c_1)|^2 = |\Gamma(1+\omega/2\pi T_H)|^2 =
(\omega/2T_H)/\sinh(\omega/2T_H)$. However, the nature of the `ingoing' distribution
functions depend on the value of $L$ that one considers. In particular, 
on substituting
the values of $a$ and $b$ from (\ref{a1}) in ${\cal {P}}$, we find that the the
gamma-functions in the numerator correspond to Fermi 
distributions for odd-$L$ and bose
distributions for even-$L$. Thus, depending on the partial wave, the vector particle
can be thought of arising out of the interactions of two bosons or two 
fermions. 

The greybody factor or the absorption coefficient of the
black hole is determined by multiplying by the plane wave
factor as:
\begin{eqnarray}
\sigma_{\mbox{abs}}&=& \frac{2 L N^{L+2}}{(L!)^4 2^{2L}}
 \frac{\omega^{2L}}{l^2 \rho_+ T_H (\rho_+ - \rho_-)^2}\nonumber\\
&\times&\frac{\sinh{\omega/T_H}}{\omega}\left|\Gamma(L/2+
i\omega/4\pi T_+)\Gamma(L/2 +1 + i\omega/4\pi T_-)\right|^2
\label{vgbf}
\end{eqnarray}
If we include rest of the components of
the six dimensional vector, i.e. $A_a$,
then the total GBF will involve a sum
of the individual greybody factors. The 
greybody factors due to $A_a$ 
are same as that of the scalars.
Since those terms do not contain the 
spin dependence, we ignore them. 

\section{CFT Description}
The decay rates are obtained from the
above greybody factors by multplying with 
the appropriate Planck or Fermi-Dirac
distributions. It has been known for
long that the these decay rates can be
reproduced form a CFT calculation
using appropriate conformal operators. Earlier, the dimensions of the 
CFT operators were guessed from the structure of the 
decay rates \cite{rot,gub}. However,
using the AdS/CFT
correspondence, the dimension as well as
the exact correlators with correct
normalisations can be determined using
prescriptions given in
\cite{wit1,poly,mathur,muck,wit2}. Here we rely on the
correspondence to determine the
correlators. We note that the near horizon approximation of the black holes can be 
used at most till $r \sim l$. 
Though the 
near horizon metric will receive corrections as $r$ approaches $l$,
we ignore them in this region. The
correlators are determined in Poincare
coordinates for convenience.

The Poincare coordinates are related to the BTZ coordinates by
the following relations:
$$w^{\pm}= \left(\frac{\rho^2 - \rho_+^2}{\rho^2 - \rho_-^2}\right)^{1/2}e^{2\pi T_{\pm}(t\pm\phi l)},\;\;\;\; x_0=
\left(\frac{N}{\rho^2-\rho_-^2}\right)^{1/2} e^{\pi T_+(t +
\phi l) + \pi T_-(t - \phi l)}.$$
The  metric in Poincare coordinates is:
\begin{equation}
ds^2= \frac{l^2}{x_0^2}\left( dx_0^2 + dw^+dw-\right)
\end{equation}
The Klein-Gordon equation on this background can be
written in the following form: 
\begin{equation}
\left[\partial_{x_0}^2 - \frac{1}{x_0}\partial_{x_0} +
4\partial_{+}\partial_- - \frac{L(L+2)}{x_0^2}\right]\phi=0
\end{equation}
Substituting:
\begin{equation}
\phi= \int d^2w \phi_k(x_0)e^{i \vec {k}.\vec {w}}
\label{fous}
\end{equation}
The solutions which are ingoing or
regular at the black hole horizon are:
$$\phi_k(x_0)= ax_0 K_{L+1}(kx_0)~.$$
Where, $k=4k_+k_-$ 
 and $a$
is an arbitrary constant of integration.
To determine the correlator corresponding to the
above scalar field and look at the bahavior
of the wavefunction at $r=l$, which implies
$x_0\sim r_0/l\approx 0$ in the dilute gas approximation.
The boundary of the AdS field is taken at $x_0=\epsilon$
where $\epsilon$ is infinitesimally small and set
set $\phi_k(\epsilon)=1$. The action is:
\begin{equation}
I=\frac{1}{2}\int d^2wdx_0 \frac{1}{2 x_0^3}\left[ g^{\mu
\nu}\partial_{\mu}\phi \partial_{\nu} \phi +m^2\phi^2\right]
\end{equation}
On partially integrating, the boundary term from this action
at $x_0=\epsilon$ is:
\begin{equation}
I_B= \int d^2w\frac1{2\epsilon} \lim_{x_0\rightarrow
\epsilon}\phi d_{x_0}\phi
\end{equation}
On using (\ref{fous}) in the above, and using the solutions
for $\phi_k(x_0)$, the  
action (fourier component) consists only of the boundary term
at $x_0=\epsilon$. The fourier component thus is: 
$$
\lim_{x_0\rightarrow \epsilon}\epsilon^{-1}\delta(k +k')\frac{K_{L+1}(kx_0)}{K_{L+1}(k\epsilon)}
d_{x_0}\frac{x_0K_{L+1}(kx_0)}{\epsilon K_{L+1}(k\epsilon)},$$
Using the
expansion for 
\begin{eqnarray}
K_{n}(kx_0)&= &\frac{1}{2}\sum_{k=0}^{k=\infty} (-1)^k \frac{(n-k-1)!}{
k!}\left(\frac{z}{2}\right)^{2k-n} +\nonumber\\
& &(-1)^{n+1}\sum_{k=0}^{k=\infty}\frac1{k!(n+k)!}\left(\frac{z}{2}\right)^{n+2k}\left[ \ln{\frac{z}{2}} - \frac12\Psi(k+1) - \frac12\Psi(k)\right],  
\label{bes}
\end{eqnarray}
 the expression reduces to:
$$\frac{2(L+1)}{(L+1)!L!}\left(\frac{k}{2}\right)^{2L+2} \epsilon^{2L+1}\ln\frac{k\epsilon}{2}$$
The leading non-analytic term has a
$\ln(k\epsilon)$ dependence. 
We keep the coefficient of the term with $\epsilon$ dependence
as $\epsilon^{2L +1} \ln \epsilon$ and fourier transform
to position space to get the correlator 
\begin{equation}
G^s(w,w')= 2(L+1)^2\frac{1}{|\vec{w}-\vec{w'}|^{2L+2}}
\end{equation}
 
For the fermionic correlators, we do a calculation similar to 
that done in \cite{muck,sfe}. The boundary is taken at
$\epsilon$. The action is taken as:
\begin{equation}
I= \int d^2wdx_0 \frac{1}{2x_0^3}\bar{\psi}(\sb +
L+1/2)\psi + C \int d^2w\bar{\psi}\psi
\end{equation}
Where $C$ is a constant, which gets fixed when we try to
obtain exact matching.
The solution of the equation of motion in the representation
of $\gamma$ matrices where $\gamma_0$ is diagonal, the two
components of $\psi$ are:
\begin{equation}
\psi_1 = \int d^2w e^{i{\vec {k}.\vec{w}}} a_1K_L(kx_0)\;\;\;
\psi_2=\int d^2 w e^{i\vec{ k}.\vec{w}}\frac{i{ \gamma.\vec{k}}}{k} K_{L+1}(kx_0)
\end{equation}
On specifying one of the components at $x_0=\epsilon$, the
other component also gets related to it. Using the above
solutions, and substituting in the boundary term of the
action, one fourier component is read as:
$$ \lim_{x_0\rightarrow \epsilon}\delta(k + k')\frac{{\vec k}.\gamma_{\alpha \beta}}{k}\frac{K_{L}(kx_0)}{K_{L+1}(k\epsilon)}$$
Taking 
 the expansion for $K_n$ as given in equation (\ref{bes}), and keeping the coefficient of the $ \epsilon^{2L+1}\ln \epsilon$ term, the
greens function in the fourier transformed space is read off as:
\begin{equation}
G^f(w,w')_{\alpha \beta}= (L+1)\frac{{ (\vec{w} - \vec{w'})}.\gamma_{\alpha
\beta}}{|\vec{w}-\vec{w'}|^{2(L
+2)}}
\end{equation}
Where $\alpha \beta$ stand for spinor indices. The correlator
has also been determined in \cite{sfe}.

To find out the correlators for the CFT operators
corresponding to the vectors, it is useful to employ
the methods of \cite{muck,yi}. 
We solve the vector equations in $AdS_3$ space,
in Poincare coordinates. 
The equation of motion for $A_0= \int d^2w e^{i\vec{k}.\vec{w}}A_0$ and $A_i=\int d^2 w
e^{i \vec{k}.\vec{w}} A_i$ have the forms:
\begin{equation}
d_{x_0}^2 A_0 - \frac{1}{x_0}d_{x_0}A_0 - \left( k^2 + \frac{L^2
-1}{x_0^2}\right)A_0=0
\end{equation}
\begin{equation}
d_{x_0}^2 A_{\pm} + \frac{1}{x_0} d_{x_0} A_{\pm} -\left(k^2 +
\frac{L^2}{x_0^2}\right) = \frac{2}{x_0}ik_{\pm}A_0
\end{equation}
The equation for $A_0$ is easily solved as $A_0= a_0 x_0
Z_L(kx_0)$, where $k^2= 4k_+k_-$. For $k^2>0$, $Z_m(kx_0)=K_m(kx_0)$
(modified Bessel function of second kind) and for $k^2<0$, this is 
$Z_m(kx_0)= J_m(kx_0)$.  
However, since we confine ourselves to
Euclidean metric, we choose the former
solution. The other two 
components are easily separated using equation (\ref{eps}).
The solutions for the two components are:
\begin{equation}
A_{\pm}= a_{\pm} x^0 K_{L\pm 1}(kx_0)
\end{equation}
The use of (\ref{eps}) also leads to a relation between
the constants $a_i's$, which implies, that only one can be
fixed independently by boundary condition. The other 
arbitrary constants are related to it, and hence are
determined. Thus only one component of $A_i$ can be fixed
at the boundary, and hence the classical source is actually
chiral. The ratio of constants are as follows:
\begin{equation}
a_{\pm}= a_0 \frac{k_{\pm}}{k}
\end{equation}
This obviously implies that $a_+/a_-= k_+/k_-$. Also,
the function $ A_-$
falls slower than $A_+$, and hence we 
specify $ A_-$ at the boundary. The other components then get related 
to it. The expression for the action is:
\begin{equation}
I= \int d^2w dx_0\frac{1}{2x_0^3}\left[\frac14 F_{\mu\nu}
F^{\mu\nu} + \frac12 L(L+2)A_{\mu}A^{\mu}\right]
\end{equation}
The boundary term which comes due to one partial integration
is:
\begin{equation}
I_B= \int d^2w \frac{x_0}{2}\left[A_+F_{0-} + A_-
F_{0+}\right]
\end{equation}
Using the solutions obtained above,  
one fourier component of the action evaluated at a distance 
$\epsilon$ is :
\begin{equation}
I = \delta(k + k')\epsilon\frac{K_{L+1}(k\epsilon)}{K_{L+1}(k\epsilon)}
\frac{k_-K_{L-1}(k\epsilon)}{k_+ K_{L+1}(k\epsilon)} 
\label{acv}
\end{equation}
On using the expansion of (\ref{bes}) we find:
\begin{eqnarray}
\frac{K_{L-1}}{K_{L+1}}&=&
\frac{(L-2)!\left(\frac{k\epsilon}2\right)^{1-L} + .. + \frac{(-1)^L}{(L-1)!}
\left(\frac{k\epsilon}2\right)^{L-1}\ln(\frac{k\epsilon}2)}{L!\left(\frac{k\epsilon}2\right)^{-1-L} +
..+ \frac{(-1)^{L+2}}{(L+1)!} \left(\frac{k\epsilon}2\right)^{L+1}\ln(\frac{k\epsilon}2)}\nonumber\\
&&\nonumber\\
&=& \frac{(-1)^L\ln(k\epsilon/2)}{L!}\left[\frac1{(L-1)!}\left(\frac{k\epsilon}{2}\right)^{2L} - \frac{(L-2)!}{L!(L+1)!}\left(\frac{k\epsilon}{2}\right)^{2L+2}\right]
\end{eqnarray}
Thus retaining only the leading power of $\epsilon$ in the
above and using that in (\ref{acv}), we get:
\begin{equation}
I = \epsilon^{1+2L}\ln \epsilon\frac{k_-^2 k^{2L-2}}{2^{2L-2}L!(L-1)!}\delta(k +k')
\end{equation}
The fourier transform of this yields in Poincare coordinates, the correlation
function as:
\begin{equation}
\left<O^vO'^v\right> = 
2 (L+1)\frac{(w^+ - w'^+)^2}{|\vec{w'} - \vec{w}|^{2(L +1)+ 2}}
\end{equation}

We now calculate the emission rates from the conformal field theory theory.
The quantum mechanical calculation
involves modelling the entire black hole
spacetime by a CFT at the boundary of the
near horizon geometry $r\sim l$. This is
in accordance with the AdS/CFT
correspondence, as the information about
the near horizon $BTZ \times S^3$ is
supposed to be encoded in the boundary
of the BTZ space. A plane wave is taken
to be incident on the black hole , which
couples with the operators of the CFT
in the region $r\sim l$. The emission
rate due to this excitation is calculated
using the results for ordinary stimulated emission.
The incident wave is
regarded as classical, while the CFT
operators are treated as quantum. The 
plane wave has to be expanded in
spherical waves, to get out the partial
wave components. The plane wave is expanded in terms
of spherical functions as:
\begin{equation}
e^{i k x}=  \sum_{L\geq 0}
\sqrt{2\pi^2}(L+1) \frac{e^{-\imath \omega r}}{(r\omega)^{3/2}} e^{\imath \psi}Z_{l,0}(cos
\theta)
\end{equation}
 The spherical wave, near $r=l$ goes as
$ r^n \phi_0 (\phi, t)$ (where n is an
integer depending on L). It couples to
the CFT operator as $\int d^2x \phi_0 {\cal {O}}$.

To determine the dimension of $\phi_0$
under conformal transformations, 
we look at the behaviour
of the wave function at $r\sim l$. 
The part of the wave function which goes
as $r^n$ 
comes from contribution of $J_{\nu}(\omega
r)$ which is analytic in the region we are considering. 
For the scalars, the wavefunction near $r\sim l$
goes as $r^L\phi_0$. As on the boundary, the
theory is invariant under conformal transformations,
$\phi_0$ should have a definite behavior under
transformations which can be scalings like $ds^2= f^2(r)(ds'^2)$.
Here the coordinates scale like $f(r)$ and the wavefunction scales 
like $\phi= f^Lr^L\phi'_0$. Since $\phi$ is a scalar, $\phi_0$
has to scale as $f^{-L}$. Which gives it a dimension of $-L$.
Accordingly, the coupling $\int d^2 x \phi_0 {\cal {O}}$ implies  the
dimension $\Delta_S= L+2$ for the operator $O$. This is
consistent with the correlator determined earlier.
 
For the fermions, the two components of the wave function do not
fall off in an identical manner, and the eigenstates of $\gamma^1$
(which is the chirality matrix for two dimensions) ,$\chi_1
+ \chi_2=\chi_+$ falls
slower than $\chi_1 - \chi_2=\chi_-$ as given in equation (\ref{in}). So for our purposes,
we take $\psi_+\approx 0$. For $$\chi_-=
\left(1/r\right)^{1/2 -L }\psi_0(t,\phi)$$
 and we get the fall off power
of $\chi_-$ as $\lambda= L-1/2$. Since the
fermion is a scalar under transformations
$r'=f(r)r$, $\psi_0$ has the dimension
of $L-1/2$ under this transformation, which
is like a conformal transformation in the
boundary metric. 
Hence by conformal
invariance of the term $\int d^2x \phi_0 {\cal O}$, the dimension
of $O$ is $\Delta_F= 2 + \lambda= L + 3/2$.
The operator $\psi_0$ is a spin $1/2$ object
under the group $SO(2,2)$, and hence,
$O$ is also spin-1/2, but of opposite
chirality. Hence, the left and right conformal
weights are determined as $h_- + h_+= L +3/2$
and $ h_- - h_+= 1/2$, thus $$ h_- = L/2 + 1, 
h_+ = L/2 + 1/2~.$$ This is the same as that
appears from the correlator calculation
given above. 
  
As for the vector field, it is immediately
observed, that the two separable components
at the boundary, $A_{1,2}= A_t \pm A_{\phi}/l$,
correspond to left moving and right moving 
sources in the boundary. The fall off in powers
of $r$ is different for the two components
and the case we are considering, and as seen
from earlier section, $A_1$ falls slower than $A_2$,
and hence $A_2\approx 0$. The fall off in
$A_1$ is as follows:
$$A_1 = r^L A_{0}(t,\phi)~.$$ Since under the
transformation $r\rightarrow f(r)r$, $A_1$ transforms
as a covariant vector, the dimension of
$A_{0}$, $\lambda= L-1$. Thus $A_0$ is a source
for the CFT operator $O_v$ with weight
$\Delta_v= 2 + L-1= L +1$. The left and right
weights can now be determined as $$h_- = L/2 +1,
\;\;\;\;\;\; h_+ = L/2\;.$$
All the weights determined above are same as those
predicted using group theoretic methods in \cite{lars}.

We are now ready to compute the emission rate due to 
the plane wave-CFT coupling $\int\phi_0 O$. 
If due to this interaction term, the state
in the CFT undergoes a transition, $\left|i\right>
\rightarrow \left|f\right>$, then, the transition
probability for this process is:
$$w_{fi} = |\phi_0|^2|\left<f \right|O\left|i\right>|^2\delta(\epsilon_f - \epsilon_i - \omega)$$
Where $\epsilon_f$, $\epsilon_i$ are the
energies at of the initial and final states of 
the CFT.
The above can be written as an integral over
the two coordinates of the boundary, and in case
the final state is not a unique state, we 
sum over the final states which gives:
$$T= 
\sum_f\int d^2xe^{i\omega t} \left<i|e^{i\epsilon_i t }O^{\dag}e^{-i
\epsilon_f t}|f\right>\left<f| O |i\right>|\phi_0|^2~.$$
If the initial state is  
the Poincare vacuum,
then the transition probability is:
\begin{equation}
T= \int d^2x e^{i\omega t}\left<0|
O^{\dag}(t)O(0)|0\right>|\phi_0|^2
\end{equation}
Essentially we need $G(w,w')|\phi_0|^2$ to complete the calculations.
For $\phi_0$ we use the form of the plane wave solutions at $r\sim l$
as determined in section III. However, as these have been determined in
BTZ coordinates, we use the conformal dimension of these when we 
use Poincare coordinates. In effect, $\phi_0^P= (2\pi T_+
w^+)^{h_+-1}(2\pi T_- w^-)^{h_- -1}(Nl^2)^{h_+ + {h_-}}\phi_0^{BTZ}$. An
additional power of $(Nl^2)^{h_+ + h_-}= (4\pi^2 T_- T_+l^4)^{h_+ +
h_-}$ enters, since in BTZ coordinates, we assume that the wave
function scales like $r^{h_+ + h_-}$ at the boundary and in the Poincare
coordinates $x_0^{-(h_+ + h_-)}$. 
($x_0=\sqrt{N}/r$ at the boundary, and we use $l$ to make the scalings in both the coordinates dimensionless)
Using  
the fact that $t= (1/4\pi
T_+)\ln
w^+ + (1/4\pi T_-) \ln w^-$, we get the integral in
the transmission coefficient to be:
$$I= \int dw^+ dw^- \frac{(w^+)^{i\omega/4\pi T_+ + h_+ -1} (w^-)^{i
\omega/4 \pi T_- + h_- -1}}{ (w^+ - 1)^{2h_+} (w^- -1)^{2h_-}}$$
where we take the initial $$w'^{\pm }= e^{ 2\pi T_{\pm}(t \pm \phi)}$$
at the origin of the BTZ coordinates.
The range of $w^{\pm}$ is from 0 to $\infty$.  Changing from
$w^{\pm} \rightarrow - w^{\pm}$, and using $B(x,y)=
\int_0^{\infty} dt~t^{x-1}/(1+t)^{x+y}$, the integral can be done:
$$I = \frac{1}{\Gamma(2h_+)\Gamma(2h_-)}e^{-\omega /2T_H}|\Gamma(h_+ 
+ i\omega /4\pi T_+)|^2 |\Gamma (h_- +
i\omega /4 \pi T_-)|^2$$
The emission rate is evaluated as:
$$(2\pi l^2T_+)^{2h_+ -1}(2\pi l^2
T_-)^{2h_--1}|\phi_0|^2 C I$$
Where C is a normalisation constant, which includes the plane
wave normalisation. Plugging in 
the correct normalisation for each of the
correlators, and using the appropriate $\phi_0$, the
emission rates are exactly same as the semiclassical
calculations. 
For the scalars, $\phi_0= 1/(L+1)!(\omega/2)^{L+1}$. Using this, as well as
$h_-=h_+=L/2 +1$, the relation
that $4 l^4\pi^2T_+T_-= N$,and multiplying by the
appropriate factor to get the plane wave normalisation, we get the emission rate as:
\begin{eqnarray}
 \Gamma^S_{\mbox{cft}}&= &2\pi N \frac{(N\omega^2)^L}{2^{2L}(L!)^4} \frac{exp(-\omega/2 T_H)}{\omega}\nonumber\\
&\times&\left|\Gamma\left(L/2+ 1 +i\omega/ 4 \pi T_+\right)\Gamma\left(L/2 +1 + i\omega / 4
\pi T_-\right)\right|^2 
\end{eqnarray}
A comparison with equation (\ref{sgbf}), shows that the
semiclassical calculation has been reproduced exactly.
There is an alternative derivation for the $s$-wave emission 
in \cite{sachs}. 

For the fermions, the wave is chosen to be of a given
chirality, and hence in the expression for the emission rate,
$\psi_0= \omega^{L+3/2} 2^{-(L+1)}/(L+1)!$
 with $h_-= L/2 +1, h_+= L/2 + 1/2$, the emission rate is
determined as:
\begin{eqnarray}
\Gamma^F_{\mbox{cft}}&=& \frac{ \pi (L +1)^2(L+2) (2\pi l^2
 T_-)^{L+1}(2\pi l^2 T_+)^{L}}{ 4 (L +1)!^2
\Gamma(L+1)\Gamma(L+2)}\left(\frac{\omega}{2}\right)^{2L}\nonumber\\
&\times& e^{-\omega/2T_H}\left|\Gamma\left( L/2 +1 +
i\omega/4\pi T_-\right)\left( L/2 + 1/2 +
i\omega/4\pi T_+\right)\right|^2
\end{eqnarray}
Using the expressions for the temperatures, it can be seen
that the above expression exactly matches that obtained in
equation (\ref{fgbf}) after multiplying by the 
Fermi-Dirac distribution, $1/(\exp(\omega/T_H) +1)$. 
The special case of $s$-waves for $T_- \gg T_+$ was obtained
in \cite{ohta2}. The GBF for the other
set of two component wave functions in
six dimension can be obtained by the same
procedure above, but now with $h_+$ and
$h_-$ interchanged.

For the vector coupling, we retain the component of the wave
function which falls slower as a function of $r$ at $r\sim
l$. This couples to the operators on the boundary.
Hence for the vector $\phi_0= A_t + A_{\phi}= 1/(L+1)!(\omega/2)^
{L+1}$. This along with $h_-= L/2 +1, h_+= L/2$,
yields the emission rate as:
\begin{eqnarray}
\Gamma^V_{\mbox{cft}}&= 2\pi&\left(\frac{\omega}{2}\right)^{2L}\frac{(L+1)^3}{\Gamma^2(L+2)}
 \frac{(2\pi l^2 T_+)^{L+1}(2\pi
l^2 T_-)^{L-1}}{\Gamma(L)\Gamma(L+2)}\nonumber \\
&\times&\frac{e^{-\omega/T_H}}{\omega}\left|\Gamma(L/2 +i\omega/4\pi T_+)\Gamma(L/2 +1 + i\omega/4\pi
T_-)\right|^2\nonumber \\
&=& 2\pi\left(\frac{N \omega^2}{4}\right)^{L}\frac{N L}{
(L!)^4(\rho_+ -\rho_-)^2}\nonumber \\
&\times&\frac{e^{-\omega/2T_H}}{\omega}\left|\Gamma( L/2 +
i\omega/ 4 \pi T_+)\Gamma(L/2 + 1+ i
\omega/ 4 \pi T_-)\right|^2
\end{eqnarray}
This is same as eqn (\ref{vgbf}) multiplied by the Planck
distribution with temperature $T_H$.
 Thus we see for each of the cases stated above the matching
is exactly obtained. 
It is interesting to note how the various factors conspire among themselves 
to yield this exact matching, using the AdS/CFT correspondence.

\section{Discussions}
In this paper, we have studied the emission rate for
particles for arbitrary partial waves by probing
the near horizon geometry of a 5-dimensional near extremal
black hole. We determined the greybody factors of scalars,
spinors and vector particles by solving their respective
equation of motion in the BTZ back ground and matching
them with wavefunctions obtained at large distances from the black hole. 
For fermions, the matching was non-trivial, and we solved the
equation of motion in an intermediate region;
$r\sim l$. The answers obtained for the scalars and spinors
reproduced the results obtained previously for the five dimensional black hole.
Our calculation for non-minimally coupled vector particles
is the first calculation for emission rates for the 
given configuration.

Next, we used the conformal field theory at the boundary to obtain the
quantum mechanical spontaneous emission rates. This is in the spirit of
the AdS/CFT correspondence, in which all information regarding the 
bulk degrees of freedom are entirely encoded in the degrees of freedom at the
boundary. Indeed, we used the various $2$-point functions which have been 
calculated from the AdS/CFT correspondence to find the decay rates, and the
latter perfectly matches with the semi classical Hawking radiation rates, 
{\it for all partial waves}. The asymptotic plane waves that excite the CFT near
$r \sim l$ carry non-trivial kinematical information and
influence the spontaneous emission rate. Thus our calculation shows how the AdS/CFT 
correspondence can be successfully used to predict the emission rates from 
black holes. The exact matching suggests that the thermodynamical properties
of these black holes are `holographically' encoded in the boundary CFT. 

It is to be noted that the CFT is at a finite distance from the horizon,
and the role of the horizon degrees of freedom
are not very clear, unlike the CFT determined in
\cite{carlip}. Also the thermodynamics of non-extremal
black holes like the Schwarzschild black hole remains
unaddressed, as the near horizon $BTZ \times S^3$ geometry
emerges only for near extremal black holes.

\noindent
\begin{center}
{\bf ACKNOWLEDGEMENTS}
\end{center}
A.D. would like to thank  S. Kalyanarama, P. Majumdar, B.
Sathiapalan, V. Suneeta and E. Witten for useful discussions. S.D. would like to 
thank M. Reisenberger for discussions. The work of S.D. was supported by 
NSF grant NSF-PHY-9514240.

\end{document}